\documentclass{aa}

\usepackage{graphicx}
\usepackage{natbib}
\usepackage[english]{babel}  
\usepackage{txfonts}
\usepackage{aalongtable,lscape}

\begin{document}

\title{Optical spectroscopy of X-ray sources in the \\
Taurus molecular cloud: discovery of ten new \\
pre-main sequence stars\thanks{Based on data collected with the Italian Telescopio Nazionale Galileo (TNG) operated on the island of La Palma by the Centro Galileo Galilei of INAF (Istituto Nazionale di Astrofisica) at the Spanish Observatorio del Roque del los Muchachos of the Instituto de Astrof{\`\i}sica de Canarias.}}

\author{L. Scelsi\inst{1} \and G. Sacco\inst{3,1} \and L. Affer\inst{1,2} \and C. Argiroffi\inst{2} \and I. Pillitteri\inst{2} \and A. Maggio\inst{1} \and G. Micela\inst{1}}

\offprints{L. Scelsi, \email{scelsi@astropa.unipa.it}}

\institute{INAF - Osservatorio Astronomico di Palermo, Piazza del Parlamento 1, I-90134 Palermo, Italy
\and
Dipartimento di Scienze Fisiche ed Astronomiche, Sezione di Astronomia, Universit\`a di Palermo, Piazza del Parlamento 1, I-90134 Palermo, Italy
\and
Consorzio COMETA, via S. Sofia 64, I-95123 Catania, Italy}

\date{Received, accepted}

\authorrunning{L. Scelsi et al.}
\titlerunning{New pre-main sequence stars in the Taurus molecular cloud}

\abstract
{}
{We have analyzed optical spectra of 25 X-ray sources identified as
potential new members of the \object{Taurus molecular cloud} (TMC), in order to
confirm their membership in this star-forming region.}
{Fifty-seven candidate members were previously selected among the X-ray sources
in the XEST survey, having a 2MASS counterpart compatible with a pre-main
sequence star based on color-magnitude and color-color diagrams.
We obtained high-resolution optical spectra for 7 of these candidates with the 
SARG spectrograph at the TNG telescope, which were used to search for lithium
absorption and to measure the H$\alpha$ line and the radial and rotational 
velocities. Then, 18 low-resolution optical spectra obtained with the instrument
DOLORES for other candidate members were used for spectral classification, for
H$\alpha$ measurements, and to assess membership together with IR color-color 
and color-magnitude diagrams and additional information from the X-ray data.}
{We found that 3 sources show lithium absorption, with equivalent widths (EWs)
of $\sim 500$\,m\AA, broad spectral line profiles, indicating rotational
velocities of $\sim 20-40$\,km\,s$^{-1}$, radial velocities consistent with
those for known members, and H$\alpha$ emission. Two of them are classified as
new weak-lined T~Tauri stars, while the EW ($\sim -9$\,\AA\ ) of the H$\alpha$
line and its broad asymmetric profile clearly indicate that the third star
(\object{XEST-26-062}) is a classical T~Tauri star. Fourteen sources observed
with DOLORES are M-type stars. Fifteen sources show H$\alpha$ emission. Six of
them have spectra that indicate surface gravity lower than in main sequence
stars, and their de-reddened positions in IR color-magnitude diagrams are
consistent with their derived spectral type and with pre-main sequence models at
the distance of the TMC. The K-type star \object{XEST-11-078} is confirmed as a
new member on the basis of the strength of the H$\alpha$ emission line. Overall,
we confirm membership to the TMC for 10 out of 25 X-ray sources observed in the
optical. Three sources remain uncertain.}
{}

\keywords{Stars: pre-main sequence -- Stars: formation -- Galaxy: open clusters and associations: Individual: Name: Taurus Molecular Cloud --  Techniques: spectroscopic}

\maketitle

\section{Introduction}
\label{intro}

The formation of stars and planets and the evolution of stellar properties
during the early stages of their lives (internal structure, angular momentum, 
magnetic activity, and X-ray emission in low-mass stars) is one of the most
intriguing problems in astrophysics. The fragmentation of large clouds and the
subsequent gravitational collapse of molecular cores is a stochastic process
that leads to a population of stellar and sub-stellar objects with a variety of
masses. The details of these physical mechanisms are still not clear;
turbulence and density fluctuations appear to be crucial
\citep{Klessen2001,Goodwin2004b}, but other factors probably play important
roles, such as the presence of magnetic fields \citep{Padoan1999} or the
shock waves propagating from supernovae explosions that may determine the
fragmentation of clouds \citep{Truelove1997}. 

Theoretical models of cloud fragmentation and core collapse should reproduce
the \emph{initial mass function} (IMF) of a star-forming region (SFR), that is
the mass distribution of a stellar group just born. Significant differences
between the IMFs derived for the \object{Taurus-Auriga SFR}
\citep{Luhman2003Toro}, which hosts a distributed mode of star formation, and
for much denser regions containing massive stars, such as \object{Orion} and
\object{IC 348}  \citep[][ respectively]{Muench2002,Luhman2003IC348}, have
suggested that the star formation process may be affected by the physical
conditions of the environment where stars form. However, previous studies of the
TMC may not be complete at low masses, while the most recent surveys
\citep{Briceno2002,Luhman2003Toro,Luhman2004,Guieu2006,Luhman2006b} have
allowed discovery of new low-mass stars and brown dwarfs, but were focused on
limited portions of the region and are not spatially complete. In fact, studies
of the TMC are also complicated by its large extension in the sky 
($\sim 100$ square degrees, also due to its distance of 140\,pc), and require 
long observational campaigns with sufficiently deep exposures. More complete 
studies of the Taurus population are therefore needed to assess the IMF shape 
with greater confidence, especially at the very low-mass end. 

We started a search for new members of the Taurus-Auriga SFR 
\citep{ScelsiXEST2006}, based on X-ray data from the \emph{{\rm XMM}-Newton
Extended Survey of the Taurus Molecular Cloud} \citep[XEST, ][]{GuedelXEST2006}
and on near-infrared data from the 2MASS point source catalog
\citep{Skrutskie2006}. This is a different approach with respect to the above 
mentioned searches for new Taurus members, where candidates were selected based
on optical and IR data. Since intense X-ray emission \citep[10 to $10^4$ times 
the solar level, see, e.g., ][]{Feigelson1999,Stelzer2001,Ozawa2005} is a
ubiquitous characteristic of pre-main sequence solar-type stars and thanks to
the relatively low interstellar absorption at these wavelengths, X-ray
observations are particularly efficient at detecting the population of young
objects, including very low-mass stars that are faint in the optical bands, and
can therefore serve as a complement to optical/IR searches. Moreover, since 
non-accreting ``weak-lined'' T~Tauri stars (WTTSs) are generally more luminous 
in X rays \citep[e.g. ][]{BriggsXEST2006} and less absorbed
than classical T~Tauri stars (CTTSs), which are still surrounded by a thick
circumstellar accretion disk, the former are expected to be selected more
efficiently in X-ray surveys. This is particularly important, because it helps
to reduce possible biases introduced by optical/IR surveys, which may favour the
detection of CTTSs owing to their strong H${\alpha}$ emission and IR excess, and
hence allows us to properly address fundamental issues such as the
estimate of disk lifetimes, with important consequences on our understanding of
the evolution of the angular momentum during the earlier phases of stellar
life and in the formation of planetary systems. 

In this paper we present the results of the analysis of follow-up optical
spectroscopy obtained at the Telescopio Nazionale Galileo (TNG) for 25 out of 57
candidate Taurus members selected in the previous work by
\citet{ScelsiXEST2006}. The paper is organized as follows. Section \ref{sel}
summarizes the main information about the XEST survey and the X-ray/IR based
selection of potential new members; in Sect. \ref{obs} we
describe the follow-up TNG observations of the optically brightest candidates.
In Sect. \ref{opt_spec} we present the results of the optical spectroscopy, 
while in Sect. \ref{IRdiagr} we determine the stellar properties. Finally, we
discuss our results in Sect. \ref{dis}.

\section{X-ray/IR selection of new TMC candidates}
\label{sel}

\citet{ScelsiXEST2006} searched for TMC candidate members among the X-ray
sources detected in the fields of the XEST survey \citep{GuedelXEST2006}. This
large project consists of 27 XMM-\emph{Newton} observations of the Taurus
molecular cloud covering the densest concentrations of molecular gas in this
region, as traced by CO emission. The exposure times are typically  
$\sim 30$\ ks each, although a few fields have longer exposures (in the range
$\sim 45-120$\ ks), while others were severely contaminated by a high level of
background radiation that reduced the useful exposure time significantly. The
total surveyed area covers about 5 square degrees of the TMC and contains 
$\sim 150$ members, that is, about half its known total population. 

Candidate members were selected among the X-ray sources with an IR counterpart
in the 2MASS catalog and not previously identified with known objects. The X-ray
sources without an IR counterpart were not considered, since their number and
global X-ray properties are consistent with their nature likely being an
extragalactic one, although some heavily absorbed Taurus members might be found
among them. We refer the reader to \citet{ScelsiXEST2006} for details about the
selection procedure, based on the consistency of the 2MASS counterparts with PMS
isochrones by \citet{Baraffe1998} and \citet{Siess2000}, calculated for the
distance of the TMC (140\,pc), in IR color-color (J-H vs. H-K) and
color-magnitude (J vs. J-K and H vs. H-K) diagrams.
Among the 57 candidates selected in this way, we identified 12 candidates with a
higher probability of membership based on their X-ray properties, namely high
plasma temperatures and/or presence of strong flares in their light curves, both
characteristic of young stars.

The candidate member \object{XEST-08-003} was also identified as a potential TMC
member by \citet{Jones1979}, who found its proper motion consistent with the
average of the Taurus members, while \object{XEST-26-062} was noticed by
\citet{Strom1976} as an IR source in the cloud \object{L1517}.

\section{Optical sample and observations} 
\label{obs}

\begin{figure}[t!]
\begin{center}
\scalebox{0.56}{
\includegraphics{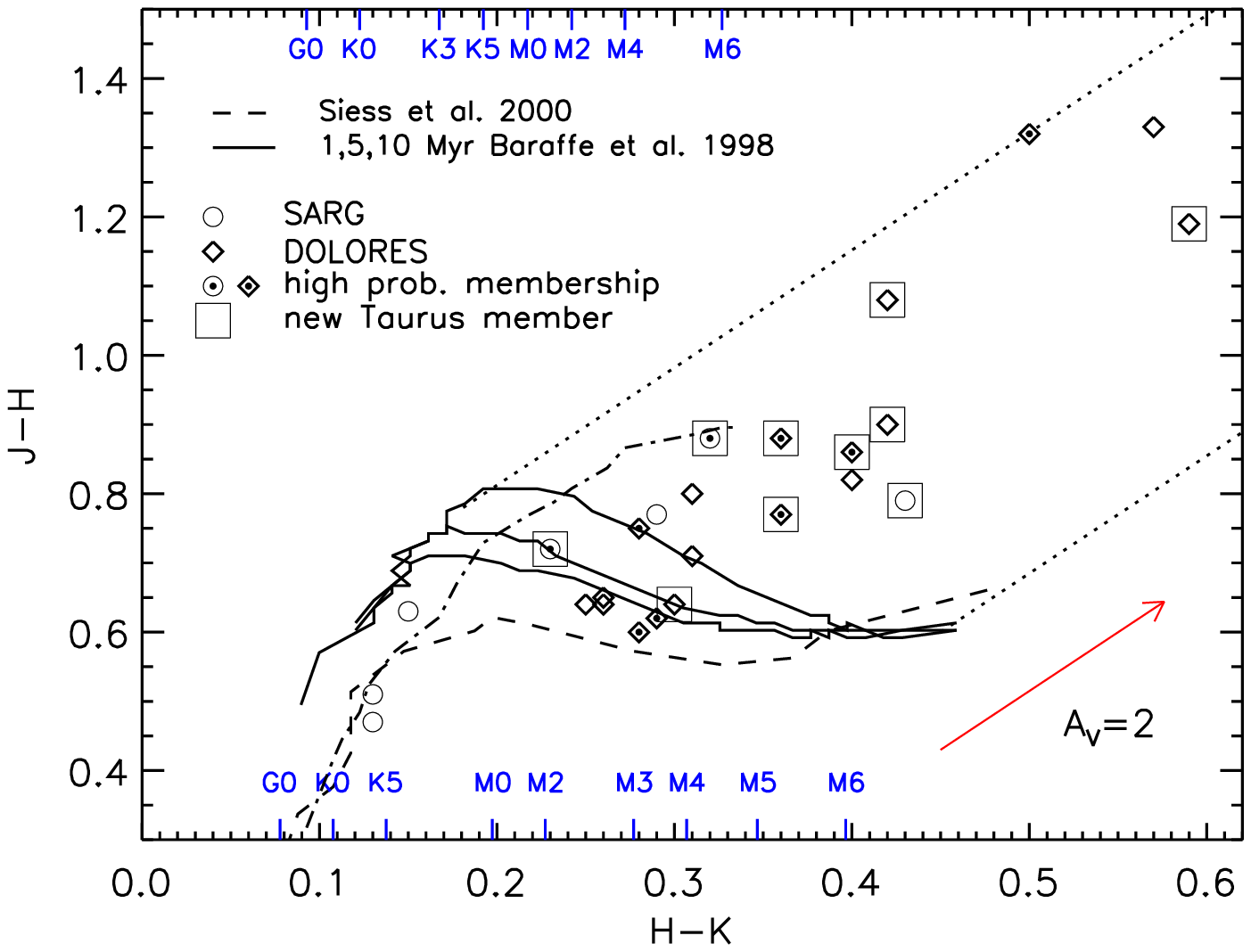}}
\scalebox{0.56}{
\includegraphics{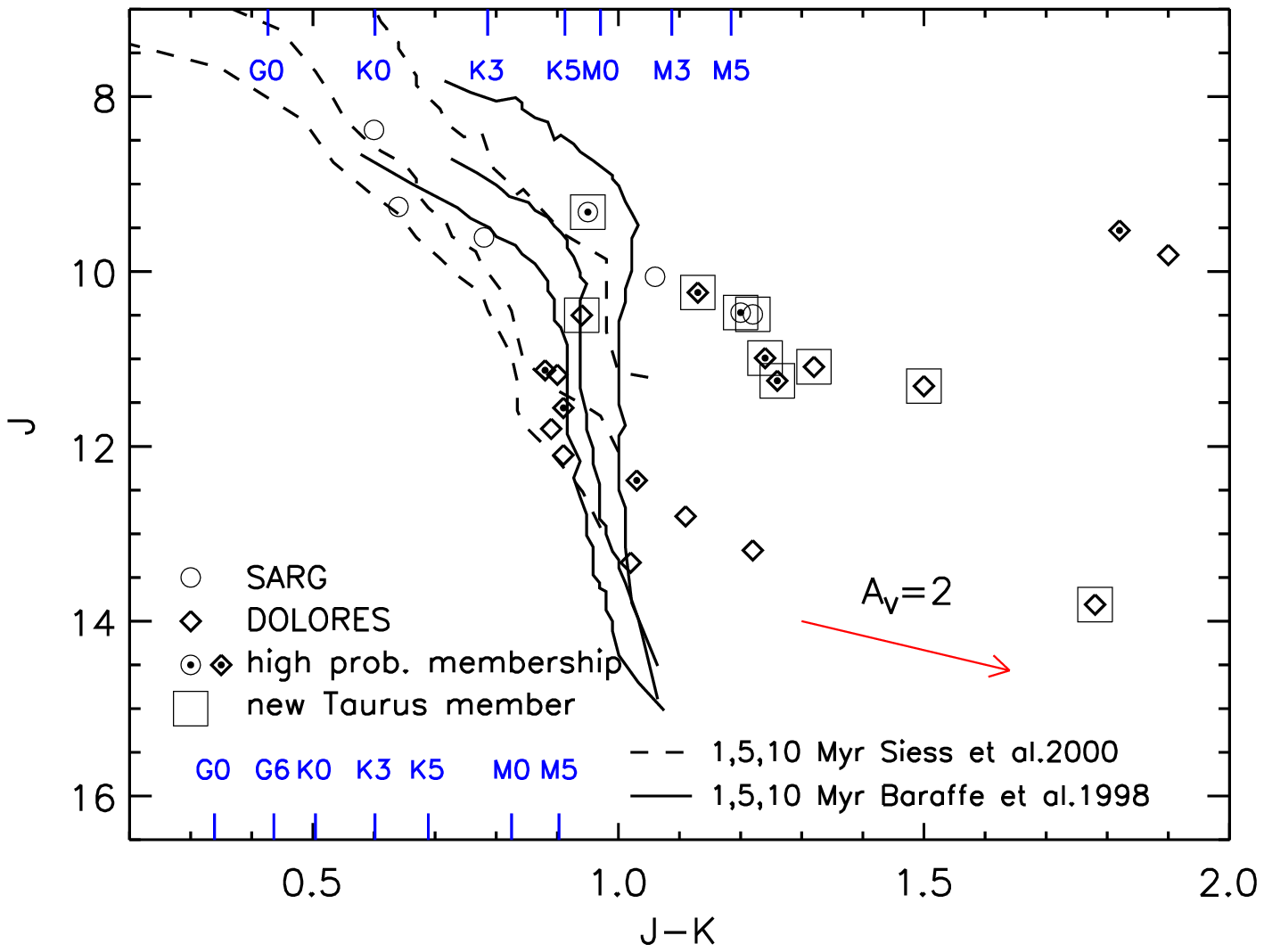}}
\caption{Color-color and color-magnitude IR diagrams for the 25 candidates 
studied in this work. Circles mark stars observed with SARG, diamonds are stars
observed with LRS. A small dot inside the symbol indicates a higher probability
of membership (see text). New members confirmed from this work are marked with
a square. Isochrones by \citet{Baraffe1998} and \citet{Siess2000}, calculated
for 1, 5, and 10\,Myr and at the distance of 140\,pc, are shown as solid and
dashed lines, respectively. In the color-color diagram, we report only the
10\,Myr isochrone by Siess et al., because all isochrones are practically
coincident. In the same plot, the locus of giant stars is shown as a dot-dashed
line, while the two dotted lines mark the region where stars without IR excess
are expected to lie. Colors of MS stars \citep{Kenyon1995} and giant stars
\citep{Bessell1988} for various spectral types are indicated, respectively, at
the bottom and top of each diagram.}
\label{fig:xest_selez}
\end{center}
\end{figure}

We chose the candidates to be observed at the TNG after
searching for optical counterparts in the USNO catalog. We selected 38 sources
with magnitude in the R band brighter than 17.5, which could be observed in a
reasonable amount of time (4 nights in total) with relatively high 
($\sim 50-80$)
signal-to-noise ratios. However, bad weather and seeing conditions during some
nights restricted the observed stars to 27. The 8 brightest stars 
(R$\simeq 10.2-13.4$) were observed at high spectral resolution with SARG, while
the remaining 19 stars (R$\simeq 13.8-17.5$) at low resolution with DOLORES/LRS.
The SARG spectrum of XEST-06-045 and the DOLORES
spectrum of XEST-26-135 turned out to be too weak and were not analyzed. The 25
sample stars considered in this work are listed in Table \ref{tab:lista},
together with relevant information, while Fig. \ref{fig:xest_selez} shows their
positions in the J-H vs. H-K and J vs. J-K diagrams.

The observations of the candidates listed in Table \ref{tab:lista} were
performed on the nights of 2007 January 20-23 with the instruments on the 3.6\,m
Telescopio Nazionale Galileo at the La Roque de los Muchachos Observatory. The
high-resolution spectrograph SARG was operated with the 61\,\AA~mm$^{-1}$ CD3
grism ($\lambda_{\rm blaze}=5890$\,\AA\ ), which produced wavelength
coverage from 4620 to 7920~\AA, with a yellow filter and 1.6'' slit providing
spectral resolution $R \sim 29000$. Exposure times ranged from 900 up to 5400\,s
for the 3 faintest sources observed with this instrument. Data reduction of
these echelle spectra (bias subtraction, flat fielding, extraction of
background-subtracted spectra, and wavelength calibration) was performed with
the tasks within the IRAF\footnote{{\tt IRAF} (Image Reduction and Analysis
Facility) is distributed by the National Optical Astronomy Observatories,
operated by the Association of Universities for Research in Astronomy, Inc., 
under cooperative agreement with the National Science Foundation.} software
system. The low-resolution spectra were obtained with DOLORES (LRS) configured
in long-slit spectroscopy mode; we used the VHR-V grism and 1'' slit. This setup
provides wavelength coverage from 4650 to 6800~\AA\ and spectral resolution of
3.8\,\AA\ ($R \sim 1500$). Exposure times ranged from 600 to 5400\,s. The LRS
spectra were reduced using both IRAF tasks and procedures written in IDL. 
These spectra were not corrected for the instrumental response and
airmass.
\begin{center}
\begin{table*}[t!] 
\caption{List of TMC candidates observed at the TNG.}
\begin{center}
\footnotesize
\begin{tabular}{lcclccccccc}\hline\hline
XEST id\hspace{1mm}$^{a}$ & RA\hspace{1mm}$^{b}$ & DEC\hspace{1mm}$^{b}$ & X-ray CR\hspace{1mm}$^{c}$ & var\hspace{1mm}$^{d}$ & desig. 2MASS &  R\hspace{1mm}$^{e}$  & J\hspace{1mm}$^{f}$ & H\hspace{1mm}$^{f}$ & K\hspace{1mm}$^{f}$ & instr.  \\ 
        & h\ \ m\ \ s  & deg\ \ $'$\ \ $''$  &    cts\ ks$^{-1}$      &     &               &  mag  &  mag  &  mag  &  mag &    \\ \hline
27-022  & 3 53 37.30 & +32 4 58.3 &  $7.9\pm 0.8$ &   & 03533730+3204582  & 10.2 & 8.38 &   7.91  &   7.78 & SARG \\  
21-059  & 4 22 25.60 & +28 12 33.4 & $8.9\pm 0.8$ &   & 04222559+2812332  & 10.8 & 9.26 &   8.75  &   8.62 & SARG \\  
11-035  & 4 21 43.73 & +26 47 22.6 & $2.1\pm 0.5$ &   & 04214372+2647225  & 11.1 & 9.61 &   8.98  &   8.83 & SARG \\  
{\bf 09-042} & 4 35 58.98 & +22 38 34.8 & $269\pm 3$ & y & 04355892+2238353  & 11.3 & 9.32 &   8.60  &   8.37 & SARG \\  
04-060  & 4 33 55.61 & +24 25 1.9 &  $1.3\pm 0.3$ &   & 04335562+2425016  & 12.8 & 10.06 &   9.29  &   9.00 & SARG \\  
26-062  & 4 55 55.99 & +30 36 21.9 & $9.5\pm 0.4$ & y & 04555605+3036209  & 13.4 & 10.47 &   9.66  &   9.27 & SARG \\  
{\bf 08-003} & 4 34 56.91 & +22 58 35.9 & $331\pm 6$\hspace{1mm}$^{g}$ & y & 04345693+2258358  & 13.4 & 10.47 &   9.59  &   9.27 & SARG \\ 
18-059  & 4 34 33.21 & +26 2 40.7  & $6.7\pm 1.1$ &   & 04343322+2602403  & 13.8 & 11.18 &  10.54  &  10.28 & LRS \\  
16-045  & 4 20 39.19 & +27 17 32.1 & $30.4\pm 2.4$ &  & 04203918+2717317  & 14.4 & 10.50 &   9.86  &   9.56 & LRS \\  
{\bf 20-071} & 4 14 52.31 & +28 6 0.3 & $159\pm 3$ &  & 04145234+2805598  & 14.4 & 9.53 &   8.21  &   7.71 & LRS \\  
06-041  & 4 04 24.48 & +26 11 12.1 & $17.1\pm 1.1$ &  & 04042449+2611119  & 14.7 & 12.10 &  11.45  &  11.19 & LRS \\  
{\bf 17-059} & 4 33 52.50 & +22 56 27.9 & $57.4\pm 1.9$ &   & 04335252+2256269  & 14.7 & 10.24 &   9.47  &   9.11 & LRS \\  
{\bf 08-014} & 4 35 13.18 & +22 59 20.5 & $62.9\pm 1.9$ & y & 04351316+2259205  & 14.9 & 11.13 &  10.53  &  10.25 & LRS \\  
{\bf 15-034} & 4 29 36.24 & +26 34 23.5 & $24.4\pm 1.0$ & y & 04293623+2634238  & 15.0 & 11.56 &  10.94  &  10.65 & LRS \\  
05-027  & 4 40 3.62 & +25 53 54.9 & $109\pm 5$ &  & 04400363+2553547  & 15.2 & 9.81 &   8.48  &   7.91 & LRS \\  
15-075  & 4 30 17.02 & +26 22 26.5 & $24.6\pm 2.5$ &  & 04301702+2622264  & 15.5 & 11.80 &  11.16  &  10.91 & LRS \\  
{\bf 08-049}  & 4 35 52.87 & +22 50 58.5 & $118\pm 2$ & y & 04355286+2250585  & 15.5 & 10.99 &  10.11  &   9.75 & LRS \\  
08-047  & 4 35 52.12 & +22 55 3.6 & $5.8\pm 0.6$ &   & 04355209+2255039  & 15.7 & 11.31 &  10.23  &   9.81 & LRS \\  
11-087  & 4 22 24.03 & +26 46 26.3 & $13.4\pm 1.1$ &  & 04222404+2646258  & 15.8 & 11.09 &  10.19  &   9.77 & LRS \\  
19-002  & 4 31 46.35 & +25 58 40.6 & $3.1\pm 0.9$ &   & 04314634+2558404  & 16.0 & 13.33 &  12.62  &  12.31  & LRS \\  
{\bf 27-084} & 3 54 10.61 & +31 48 58.2 & $18.0\pm 0.7$\hspace{1mm}$^{g}$ & y & 03541064+3148573  & 16.1 & 12.39 &  11.64  &  11.36 & LRS \\  
{\bf 08-033} & 4 35 42.03 & +22 52 22.5 & $27.2\pm 0.9$ &  & 04354203+2252226  & 16.3 & 11.25 &  10.39  &   9.99 & LRS \\  
22-111  & 4 32 26.88 & +18 18 23.1 & $9.9\pm 1.3$ & y & 04322689+1818230  & 16.4 & 12.80 &  12.00  &  11.69 & LRS \\  
11-078  & 4 22 15.68 & +26 57 6.4 & $9.4\pm 0.5$ &  y & 04221568+2657060  & 17.1 & 13.81 &  12.62  &  12.03 & LRS \\  
12-012  & 4 34 51.64 & +24 4 43.0 & $3.3\pm 0.8$ &    & 04345164+2404426  & 17.5 & 13.19 &  12.37  &  11.97 & LRS \\  \hline
\multicolumn{10}{l}{Note: sources in bold face have a higher probability of
membership on the basis of X-ray light curve and/or spectral } \\
\multicolumn{10}{l}{analysis \citep[see ][]{ScelsiXEST2006}.} \\
\multicolumn{10}{l}{$^a$ Name of the X-ray source in the XEST catalog (the first 2 digits mark the field of the survey).} \\ 
\multicolumn{10}{l}{$^b$ Coordinates of the optical counterpart in the USNO catalog.} \\ 
\multicolumn{10}{l}{$^c$ Equivalent on-axis count-rate for the PN in the $0.5-7.3$\,keV band, averaged over the entire observation.} \\ 
\multicolumn{10}{l}{$^d$ X-ray source variable according to the maximum likelihood algorithm described in \citet{StelzerXEST2006}.} \\ 
\multicolumn{10}{l}{$^e$ Magnitude in the R band reported in the USNO catalog.} \\ 
\multicolumn{10}{l}{$^f$ Infrared magnitudes reported in the 2MASS catalog.} \\ 
\multicolumn{10}{l}{$^g$ Quiescent count-rates (cts\,ks$^{-1}$) for sources with
strong flares: $180\pm 40$ (\object{XEST-08-003}), $15.0\pm 4.3$ (\object{XEST-27-084}).} \\
\end{tabular}
\normalsize
\end{center}
\label{tab:lista}
\end{table*}
\end{center}

\section{Results}
\label{opt_spec}

\subsection{High-resolution spectra}
\label{sarg}

\begin{figure*}[t!]
\begin{center}
\scalebox{0.83}{
\includegraphics{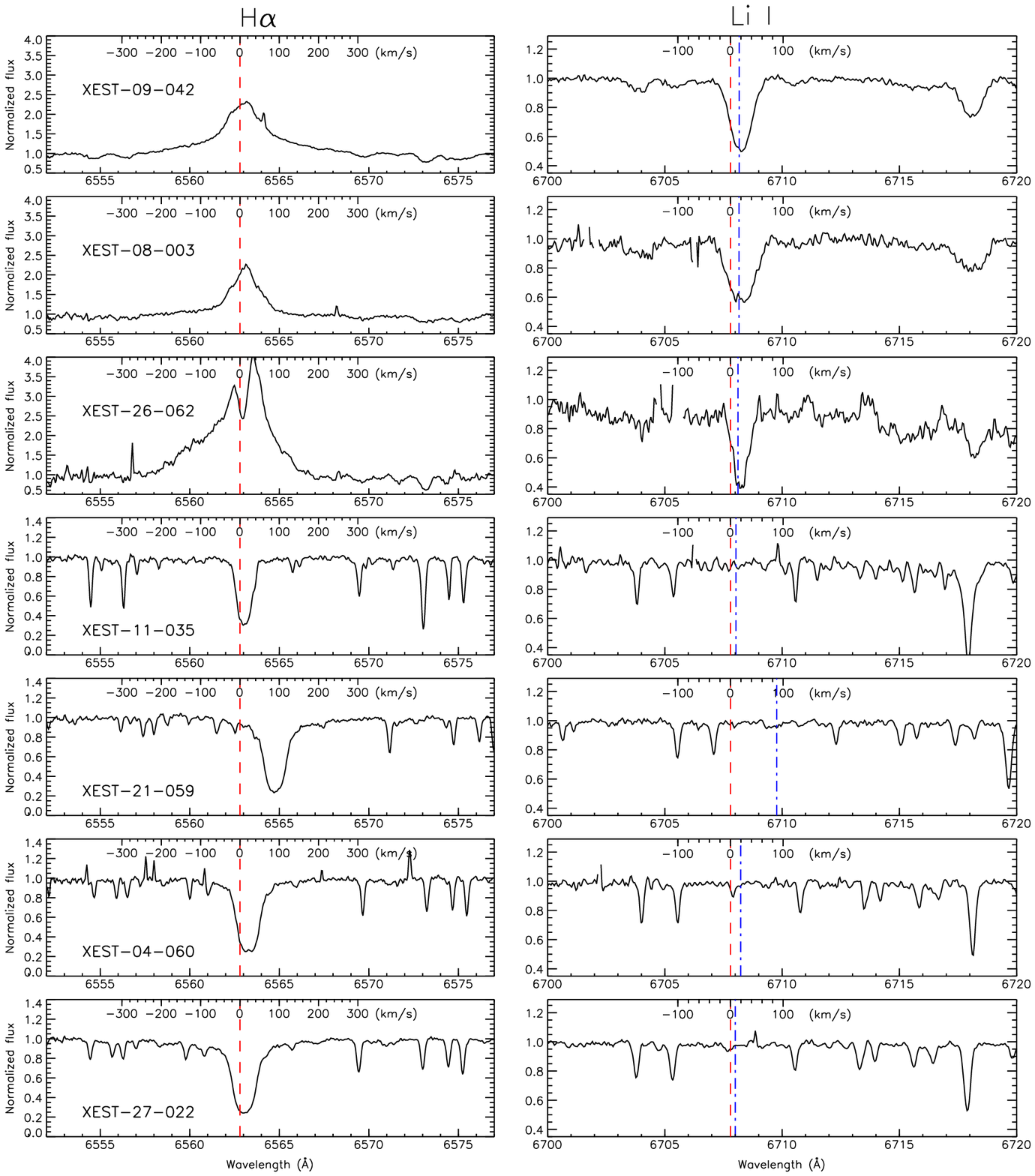}}
\caption{High-resolution spectral regions (normalized to the continuum) centered
around the H$\alpha$ (\emph{left panels}) and Li {\sc i}\,6708\AA\ (\emph{right
panels}) lines. The red dashed vertical line in each plot marks the nominal
wavelength of the spectral line; the blue dot-dashed lines in the right panels
show the expected positions in the stellar spectrum of the Li line, based
on the radial velocity shift. A velocity scale is reported on top of each panel.
All prominent and narrow emission features in the plotted spectra (including
that in the right side of the H$\alpha$ profile of XEST-09-042) are due to
cosmic rays or hot pixels.}
\label{fig:Li_Halpha}
\end{center}
\end{figure*}

\begin{center}
\begin{table*}
\caption{Measures of the equivalent widths of the Li {\sc i} 6708\AA\ and
H$\alpha$ lines and of radial and rotational velocities for the 7 stars observed
with SARG. }
\begin{center}
\begin{tabular}{lccccc}\hline\hline
star        &     Li EW    & H$\alpha$ EW & $v_{\rm rad}$ & $v_{\rm rot}\,\sin i$ & Spectral \\ 
            &      \AA\    &     \AA\     &  km\,s$^{-1}$ &  km\,s$^{-1}$ & classification     \\ \hline
\multicolumn{6}{c}{new TMC members} \\
\object{XEST-09-042} & $0.53\pm 0.03$ &  $-3.2\pm 0.5$ &  $16.7\pm 1.0$   & $31.3^{+2.8}_{-2.1}$ & K5-7 V/IV \\ 
\object{XEST-08-003} & $0.56\pm 0.05$ & $-2.5^{+0.4}_{-0.7}$ &  $16.2\pm 1.5$   & $44^{+4}_{-3}$ & K5-7 V/IV \\ 
\object{XEST-26-062}\,$^{a}$ & $0.46\pm 0.06$ &  $-9.0\pm 1.5$ &  $14.4\pm 1.1$   & $18\pm 3$ & M  \\ \hline
\multicolumn{6}{c}{non members} \\
\object{XEST-11-035} & $< 0.03$ &  $0.76^{+0.15}_{-0.06}$ &  $10.39\pm 0.25$ & $< 4$ & K5 V \\
\object{XEST-21-059} & $< 0.025$ & $1.5\pm 0.2$  &  $88.10\pm 0.23$ & $< 5$ & K1 IV/III \\ 
\object{XEST-04-060} & $< 0.025$ & $1.4\pm 0.2$  &  $19.34\pm 0.23$ & $< 5$ & K1 IV/III \\
\object{XEST-27-022} & $< 0.02$ & $1.45\pm 0.15$ &  $9.13\pm 0.19$  & $< 5$ & K1 IV/III \\ \hline
\multicolumn{6}{l}{$^a$ Absolute values of the equivalent widths of the Li
and H$\alpha$ lines are likely underestimated} \\
\multicolumn{6}{l}{\hspace{0.2cm} since veiling was not accounted for.} \\
\end{tabular}
\normalsize
\end{center}
\label{tab:misure_sarg}
\end{table*}
\end{center}
The SARG spectra of seven candidates (see Table \ref{tab:lista}) were used to
measure the strength of Li absorption and H$\alpha$ line (Fig.
\ref{fig:Li_Halpha}) and to measure radial and rotational velocities. We also 
performed spectral classification. Typical signal-to-noise ($S/N$) ratios
of the SARG spectra are in the range $60-80$ for all stars, with the exception of \object{XEST-08-003} ($S/N \sim 45$) and \object{XEST-26-062} 
($S/N \sim 35$). Table \ref{tab:misure_sarg} reports the above-mentioned
measurements for the seven stars. The lithium {\sc i}\ 6708\,\AA\ absorption
line is visible in the spectra of 3 stars (\object{XEST-09-042}, 
\object{XEST-08-003}, and \object{XEST-26-062}); the same spectra show
H$\alpha$ emission, as well as broadened absorption lines indicative of high
rotational velocities. Moreover, their radial velocity is similar to those of
known Taurus members. In the light of these results detailed in the following
Sects. \ref{litio}, \ref{H_alpha}, and \ref{rad_rot_vel}, we identify these
stars as new pre-main sequence stars of the Taurus-Auriga region. In the spectra
of the other four stars, H$\alpha$ is in absorption and all absorption features
are narrow. 
\vspace{0.5cm}
\subsubsection{Lithium absorption}
\label{litio}

The EW of the Li {\sc i}\ 6708\,\AA\ line was measured for the stars
\object{XEST-09-042}, \object{XEST-08-003}, and \object{XEST-26-062} by
numerical integration along the spectrum profile, after normalization to the
continuum. For each of these measurements, errors were estimated by repeating
the calculation with different levels of the continuum.

As shown in Sect. \ref{sarg_sptypes}, \object{XEST-09-042}, \object{XEST-08-003}, and \object{XEST-26-062} have a spectral type likely in
the range K5-M3. Lithium absorption at 6708\,\AA\ with EW of 
$\sim 500-600$\,m\AA\ (see Table \ref{tab:misure_sarg})
implies Li abundances A(Li)\,$\sim 2.7-3.2$ \citep{Soderblom1990} that, in
turn, place all these stars younger than $\sim 15$\,Myr, using the models by
\citet{Siess2000}. 

For each of the other four stars, we give in Table \ref{tab:misure_sarg} a rough
upper limit to the EW of the Li line. Since we do not see any
absorption line at the expected wavelength of the Li line (after correction
for radial velocity, see Sect. \ref{rad_rot_vel}), we fitted with a Gaussian the
weakest Fe line discernible in the spectrum, close to the expected Li line
position, and set its EW equal to the upper limit of the Li line. Hereafter, we
will consider these four stars as non members.

\subsubsection{H$\alpha$ emission}
\label{H_alpha}

\begin{figure}[t!]
\begin{center}
\scalebox{0.6}{
\includegraphics{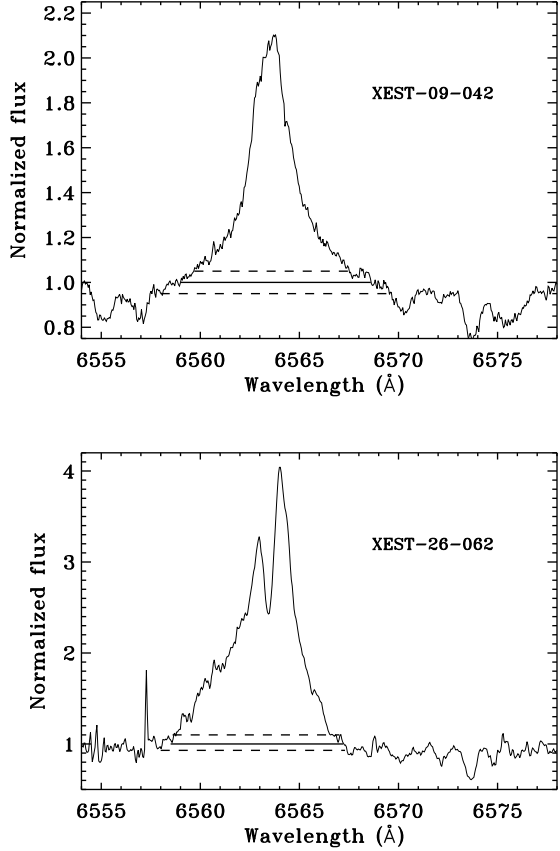}}
\caption{Measurement of the H$\alpha$ line in the spectra of
\object{XEST-09-042} and \object{XEST-26-062} normalized to the continuum. The
three horizontal lines represent the three different levels of continuum
emission assumed for the measurement of the EW (solid line) and the estimate of
the errors (dashed lines).}
\label{fig:es_Halpha_meas}
\end{center}
\end{figure}

The H$\alpha$ emission line in the spectra of \object{XEST-09-042},
\object{XEST-08-003}, and \object{XEST-26-062} (Fig. \ref{fig:es_Halpha_meas})
was measured in the same way as for the Li line, while for the four non-members
we fitted the absorption line profile with a Voigt function. We noted that the
H$\alpha$ line is partially saturated in all of these last stars, hence the
values reported in Table \ref{tab:misure_sarg} are likely to be overestimated.

In the spectrum of \object{XEST-26-062}, the profile of the H$\alpha$ emission
line is broad and asymmetric, with absorption in its central part. The full
width at 10\% of the peak is $\sim 320$\,km\,s$^{-1}$, and surely above the
limit of $\sim 270$\,km\,s$^{-1}$ that divides classical from weak-lined T~Tauri
stars according to \citet{White2003} also taking into account the broadening due
to the stellar rotation (see Table \ref{tab:misure_sarg}). These results clearly show that \object{XEST-26-062} is a CTTS. The value of EW for this star 
($\sim -9$\AA\ ) may be underestimated due to the likely presence of veiling.
Instead, the stars \object{XEST-09-042} and \object{XEST-08-003} have
EW(H$\alpha$) of $-2.5$\,\AA\ and $-3.2$\,\AA, respectively, and we have
classified them as new weak-lined T~Tauri stars.

\subsubsection{Spectral classification}
\label{sarg_sptypes}

We performed spectral classification of these 7 sources by employing the whole
red region and the $\sim 200$\,\AA\ wide region around the $5890-5896$\,\AA\ 
Na {\sc i} doublet of our SARG spectra. We compared the sodium region and the
$\sim 70$\,\AA\ wide spectral region centered on the H$\alpha$ with
high-resolution UES spectra of standard stars from \citet{Montes1998}. Moreover,
we reduced the spectral resolution of the SARG spectra to $R\sim 2400$, to
compare them with low-resolution spectra included in the library by
\citet{Sanchez2006}. Following \citet{Strassmeier1990}, the ratio of residual
intensity of Co {\sc i} 6450\,\AA\ and Ca {\sc i} 6449\,\AA\ lines allowed us to
assess the luminosity class for non-members.

Spectral classification is quite secure for the K-type field stars, for which
normalization to the continuum of the SARG spectra was straightforward, and
their spectral lines are well-resolved. By comparison with the spectra of
standard stars, errors on the spectral types were estimated to be $\pm 1$
sub-type. In contrast, broadened lines, uncertain continuum normalization,
and veiling in the case of \object{XEST-26-062}, made the classification of the
three new members more uncertain. \object{XEST-09-042} and \object{XEST-08-003}
have very similar spectra, which suggests a $mid/late$-K type, while 
\object{XEST-26-062} is most likely an $early$-M type star. Spectral types and
luminosity classes of the 7 sources are reported in Table \ref{tab:misure_sarg}.

\subsubsection{Radial and rotational velocities}
\label{rad_rot_vel}

Radial and rotational velocities for each of the 7 stars were measured 
simultaneously with the task {\sc rv} within IRAF, by cross-correlating the 
observed SARG spectrum with an adequate template spectrum. 
Template spectra at a resolution of $R=29000$ were synthesized by using
ATLAS9 \citep{Kurucz1970} model atmospheres, with temperatures and gravities
as expected for the stars analyzed here (see Sects. \ref{sarg_sptypes} and 
\ref{IRdiagr_sarg}).

The peak of the cross-correlation function was fitted with a Gaussian plus a
background level. We found that the correlation peak for the case of the new
member \object{XEST-08-003} is slightly asymmetric, which may indicate a binary
nature for this star. 

The relation between the FWHM of the peak and the stellar rotational velocity 
$v_{\rm rot}$ was calibrated by correlating the synthesized template spectrum
with artificially broadened templates for different values of stellar rotation.
We noted that the $v_{\rm rot}$--FWHM relation is almost independent of the
effective temperature and gravity of the template. Moreover, given the
resolution and $S/N$ of our SARG spectra, the measurement of rotational
velocities significantly below $\sim 10$\,km\,s$^{-1}$ is difficult. 

Table \ref{tab:misure_sarg} contains the radial and rotational velocities 
for the 7 stars observed with SARG. The radial velocities of
\object{XEST-09-042}, \object{XEST-08-003}, and \object{XEST-26-062} are in very
good agreement with that of the Taurus-Auriga association \citep[$v_{\rm rad}=16.03 \pm 6.43$\,km\,s$^{-1}$, measured by ][ based on a sample of 127 Taurus members]{Bertout2006}, and thus is another indication that these stars
belong to the TMC. Their rotational velocities are typical of other young stars 
\citep[see, for example, the tables in ][]{GuedelXEST2006}, while the 4 field
stars are slow rotators with projected rotational velocities lower than 
$\sim 4-5$\,km\,s$^{-1}$ and radial velocities not consistent with Taurus
members.

\subsection{Low-resolution spectra}
\label{dolores}

\begin{figure*}[t!]
\begin{center}
\scalebox{0.9}{
\includegraphics{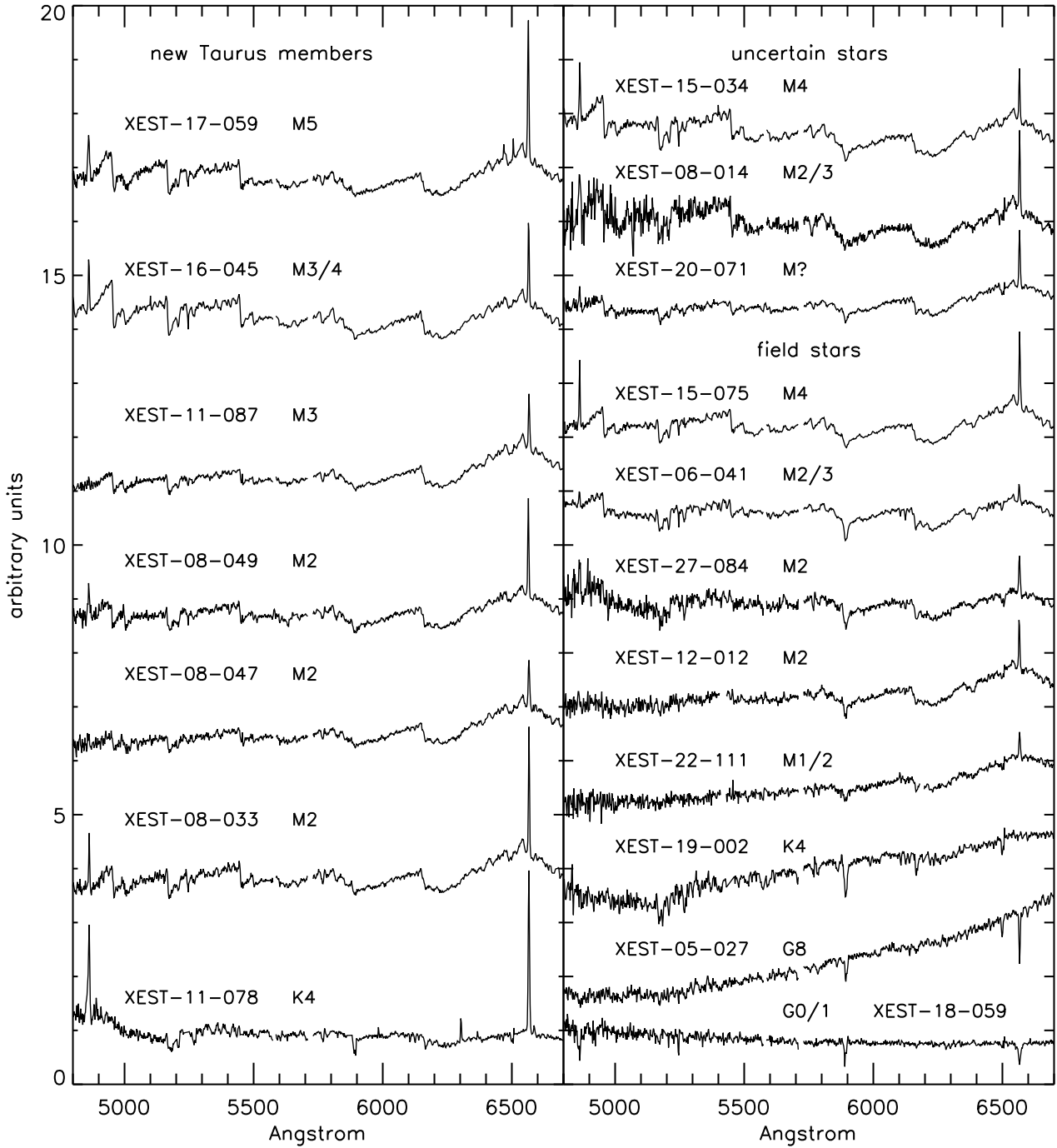}}
\caption{Low-resolution spectra of the 7 new Taurus members (\emph{left}), the
three uncertain cases (\emph{upper right}) and the field stars (\emph{lower
right}). Recall that these spectra are not corrected for the instrumental
response and air mass effects.}
\label{fig:spettri_LRS}
\end{center}
\end{figure*}

We derived spectral types for the 18 candidate members listed in Table
\ref{tab:lista} by employing low-resolution spectra obtained with DOLORES (Fig.
\ref{fig:spettri_LRS}). We also inferred the presence of H$\alpha$ emission in
15 of these stars, indicative of magnetic activity (and accretion in classical
T~Tauri stars). In Tables \ref{tab:stars_lrs_memb_unc} and
\ref{tab:stars_lrs_fs} we give a rough estimate of the H$\alpha$ equivalent
width, measured by integration of the spectral counts. Errors on the EW are in
the range $10-30\%$, due to the limited spectral resolution and to the
uncertainty on the correct continuum level. 

Spectral classification was performed through
visual comparison of our stellar spectra with those, at similar resolution, of
standard stars of various spectral types (G, K, and M) and luminosity classes
(dwarfs and giants) in the libraries of \citet{Jacoby1984} and
\citet{Sanchez2006}. In particular, we based our classification upon the
presence and strength of the following spectral features: Mg {\sc i} triplet at
$\sim 5167-5184$\,\AA\ and the Cr {\sc i} line at $\sim 5206$\,\AA; the blend of
Fe {\sc i}, Ca {\sc i}, and Ti {\sc i} lines at $\sim 5495-5515$\,\AA; the Na
{\sc i} doublet at $5890-5896$\,\AA; the Ca {\sc i} and Fe {\sc i} lines in the
range $\sim 6100-6200$\,\AA; the large blend of Ti {\sc i}, Fe, and calcium
hydride (CaH) lines in the range $\sim 6350-6410$\,\AA; the unresolved 
Fe {\sc i}, Ca {\sc i}, and Ba {\sc ii} lines around $\sim 6495$\,\AA; the
molecular bands $\sim 200$\,\AA\ wide of titanium oxide (TiO) centered around
$\sim 5950$\,\AA\ and $\sim 6250$\,\AA; and the sharp edges at $\sim 4750$\,\AA,
$\sim 4950$\,\AA, $\sim 5170$\,\AA, and $\sim 5450$\,\AA. We also used the
relative strength of the H$\alpha$ and H$\beta$ absorption lines in the spectra
of the two G-type stars of our sample. 

Low-resolution spectra of G-type stars appear relatively featureless in the
spectral range $4800-6800$\,\AA, with rather weak absorption lines from Na,
Mg, and Fe and prominent H$\alpha$ line as the main features; strong 
Na {\sc i} doublet, Mg {\sc i} triplet, and Ca {\sc i} $6100-6200$\,\AA\ lines
are the most distinguishing characteristics of K-type stars, while the cooler
atmospheres of M-type stars are easily recognized from the TiO molecular bands.
We identified 2 G-type stars and 2 K-type stars, while the remaining 14
sources are classified as M-type stars. Errors on the spectral type are
typically $\sim 2$ sub-types, small enough for membership
purposes (see Sect. \ref{IRdiagr_lrs}). As we show in Sect.
\ref{IRdiagr_lrs}, the G-type stars and one K-type star cannot be consistent
with being pre-main sequence stars, regardless of their luminosity classes.
Instead, the K-type star \object{XEST-11-078} shows H$\alpha$ emission with EW 
$\sim -16.5$\,\AA, atypical in active K-type dwarfs. In its spectrum we also
identified the [O {\sc i}]\,6300\AA\ and [N {\sc ii}]\,6583\AA\ emission lines,
typical of young accreting stars, so hereafter we consider XEST-11-078 a new
classical T~Tauri star belonging to Taurus-Auriga.

\begin{figure*}[t!]
\begin{center}
\scalebox{1.0}{
\includegraphics{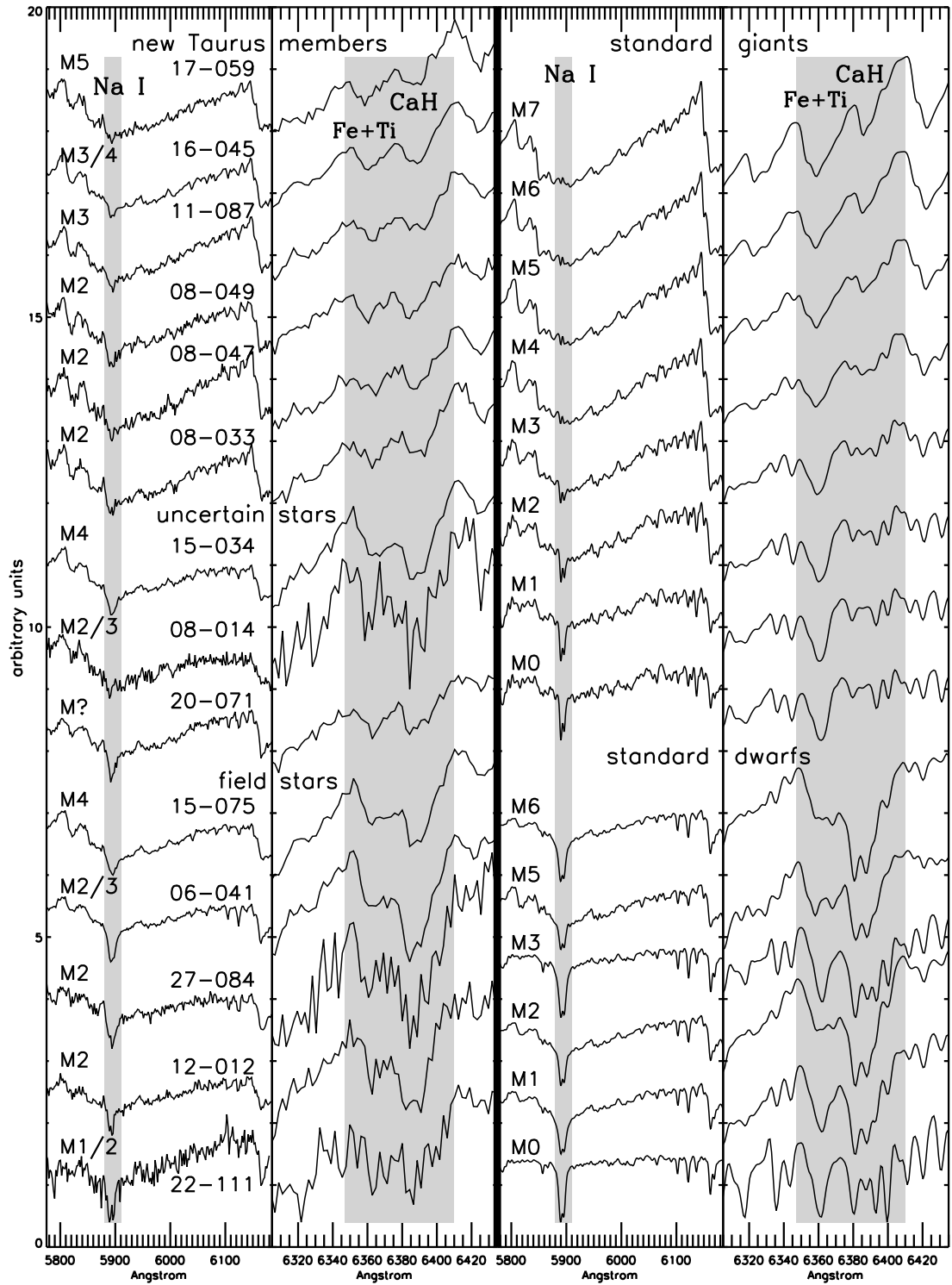}}
\caption{\emph{Left panels:} Zooms of DOLORES spectra of the 7 new Taurus
members (up), the three uncertain cases (middle) and the field stars
(bottom), showing the Na {\sc i} doublet ($\sim 5890$\,\AA) and the Fe+Ti 
($\sim 6360$\,\AA) and CaH ($\sim 6390$\,\AA) absorption features. \emph{Right
panels:} Same spectral regions for a sequence of M-type standard giants (up) and
dwarfs (bottom).}
\label{fig:Na_CaH}
\end{center}
\end{figure*}

For the M-type sources, it is crucial to assess their luminosity class, in order
to distinguish between MS or giant field stars and pre-main sequence objects.
Two well-marked spectral features in our DOLORES spectra can be reliably used
for gravity diagnostics, namely the CaH band ($\sim 6380-6390$\,\AA), to be
compared with the close absorption feature at $\sim 6360$\,\AA\ due to Ti and Fe
lines, and the sodium doublet ($5890-5896$\,\AA). In Fig. \ref{fig:Na_CaH} we
zoom in on such spectral regions for our M-type stars, and show for comparison
the same regions in the spectra of standard dwarfs and giants from the library
of \citet{Sanchez2006}, smoothed at the same resolution. Sodium absorption is
strong in standard dwarfs of all sub-types, while it weakens considerably in
giants from M2 towards later spectral types. The CaH band is stronger than the
adjacent Ti and Fe absorption feature in dwarfs later than M0; on the contrary,
it is much weaker than the Ti and Fe blend in giants of all sub-types. 

First, we observe that 5 stars labeled in Fig. \ref{fig:Na_CaH} as
``field stars'' (\object{XEST-22-111}, \object{XEST-12-012},
\object{XEST-27-084}, \object{XEST-06-041}, and \object{XEST-15-075}) show
prominent absorption from sodium and CaH, which closely resemble the spectra of
standard dwarfs. As stated in Sect. \ref{IRdiagr_lrs}, these stars are
compatible with being MS field stars. 

The 6 stars labelled  as ``new Taurus members'' (\object{XEST-08-033},
\object{XEST-08-047}, \object{XEST-08-049}, \object{XEST-11-087}, 
\object{XEST-16-045} and \object{XEST-17-059}) differ markedly from the
stars of the previous group and the standard dwarfs. In fact, sodium absorption
is much weaker and the CaH band, compared with the 6360\,\AA\ absorption
feature, shows intermediate behavior between dwarfs and giants, as one could
expect for intermediate-gravity stars like pre-main sequence objects. We stress
that both Na {\sc i} and CaH lines consistently indicate a gravity lower than in
MS stars for these stars. Such information, together with their consistent
position with pre-main sequence models in IR color-color and color-magnitude
diagrams (Sect. \ref{IRdiagr_lrs}), lead to the conclusion they are young
objects belonging to the TMC.

\object{XEST-15-034} shows signs of high gravity, while the situation is less
clear for \object{XEST-08-014} and \object{XEST-20-071} from the behavior of the
Na {\sc i} lines and CaH band. These three stars are labeled as ``uncertain'' in
the figure, for reasons also discussed in Sect. \ref{IRdiagr_lrs}.

\section{IR diagrams, membership, and stellar properties}
\label{IRdiagr}

The IR color-color and color-magnitude diagrams were employed to derive
information about the stellar properties and the interstellar absorption 
$A_{\rm V}$ and to assess membership in the case of the stars observed at low
resolution. Using the J, H, and K magnitudes from the 2MASS catalog, we plotted
the sources in the J vs. J-K, H vs. H-K, and J-H vs. H-K diagrams, together with
isochrones and the main sequence calculated for the distance of the TMC
(140\,pc). 

We assumed that the stars analyzed here have no significant IR excess from a
circumstellar disk, and therefore estimates of $A_{\rm V}$ can be derived simply
by shifting back their positions in the color-color diagram to PMS isochrones,
or to the MS or the giant locus along the reddening vector (Fig.
\ref{fig:stima_massa}). We discuss individual cases in Sects. \ref{IRdiagr_sarg}
and \ref{IRdiagr_lrs} where we know, or we suspect, this assumption is not true.

The results reported in this section were obtained using isochrones and 
evolutionary tracks by \citet{Baraffe1998}, but we also verified that our 
overall conclusions do not change substantially if we use the models by 
\citet{Siess2000}.

Distances of main sequence field stars were calculated by matching the
dereddened positions with the MS, scaled for the distance, both in the 
J vs. J-K and in the H vs. H-K diagrams, and averaging the two values, if
different. 

Masses and X-ray luminosities were also estimated. The method used to derive
masses and relevant errors for PMS stars is described in
Sects. \ref{IRdiagr_sarg} and \ref{IRdiagr_lrs}. Masses of MS stars have been 
evaluated from dereddened IR colors and the MS by \citet{Baraffe1998}. For
X-ray luminosities, we used the relation between $L_{\rm X}$, $N_{\rm H}$ and
the EPIC/PN count-rate found for known TMC members \citep{ScelsiXEST2006}.
We have verified that such a relation approximately holds
also for coronae cooler ($T\sim 0.3$\,keV) than those of young active stars, if 
$N_{\rm H}\lesssim 10^{22}$\,cm$^{-2}$, so it was adopted for 
computing X-ray luminosities of both new TMC members and MS field stars. 
For giants (or subgiants) field stars, we evaluated only the 
$L_{\rm X}/L_{\rm bol}$ ratios (using the tables of intrinsic colors 
from \citealt{Zombeck1991}).

Rough estimates of the errors on extinctions, masses, and distances of
MS field stars were calculated by propagating the errors on the 
IR colors and the spectral types.

\subsection{Stars observed with SARG}
\label{IRdiagr_sarg}

The results shown in Sect \ref{sarg} confirm membership to the
TMC of \object{XEST-09-042}, \object{XEST-08-003}, and \object{XEST-26-062}, for
which we derived here estimates of visual absorption, mass, and age. The four
remaining sources are one foreground main sequence stars and three K1 IV/III
stars behind the cloud. 

\begin{figure}[t!]
\begin{center}
\scalebox{0.5}{
\includegraphics{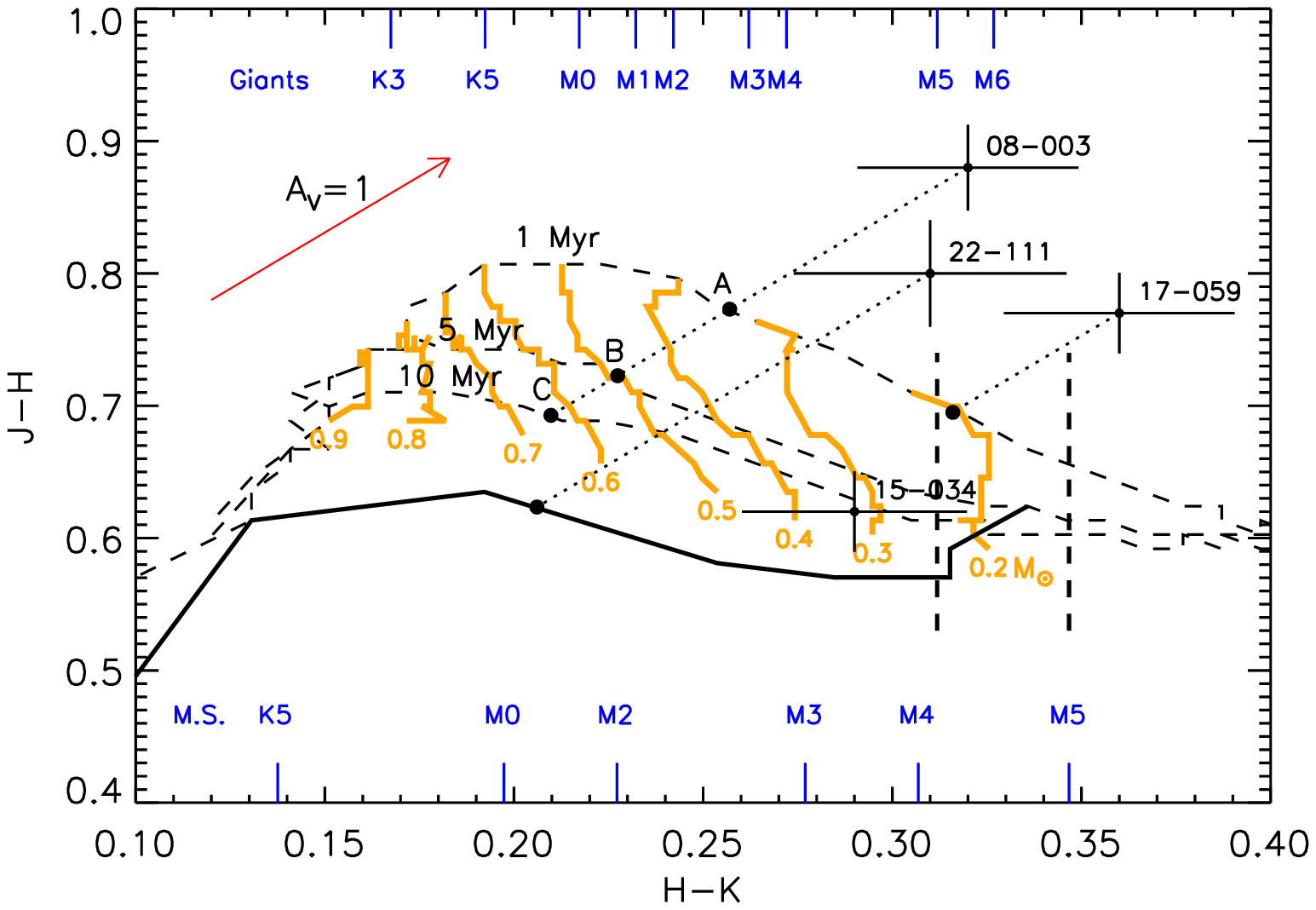}}
\scalebox{0.5}{
\includegraphics{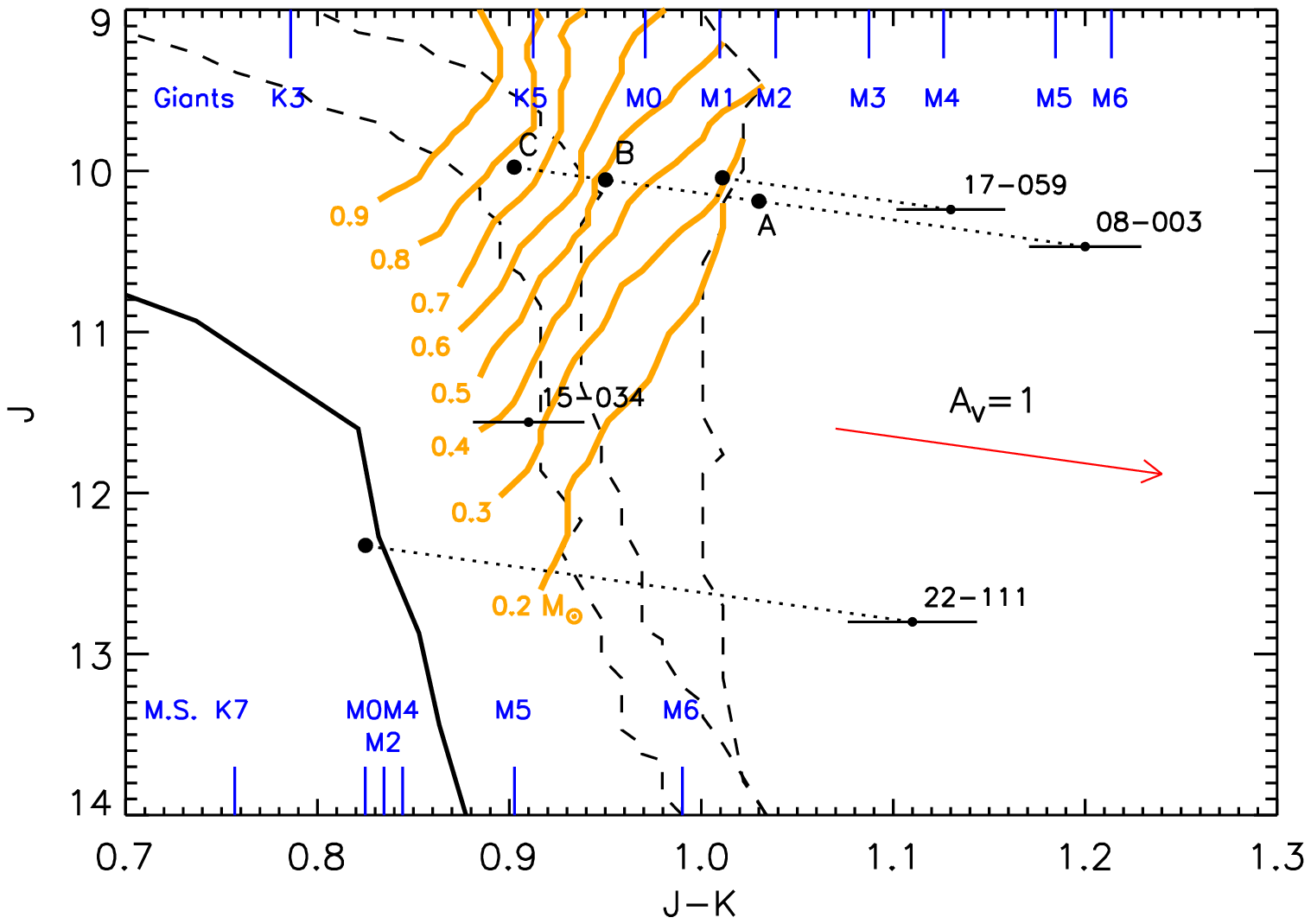}}
\scalebox{0.5}{
\includegraphics{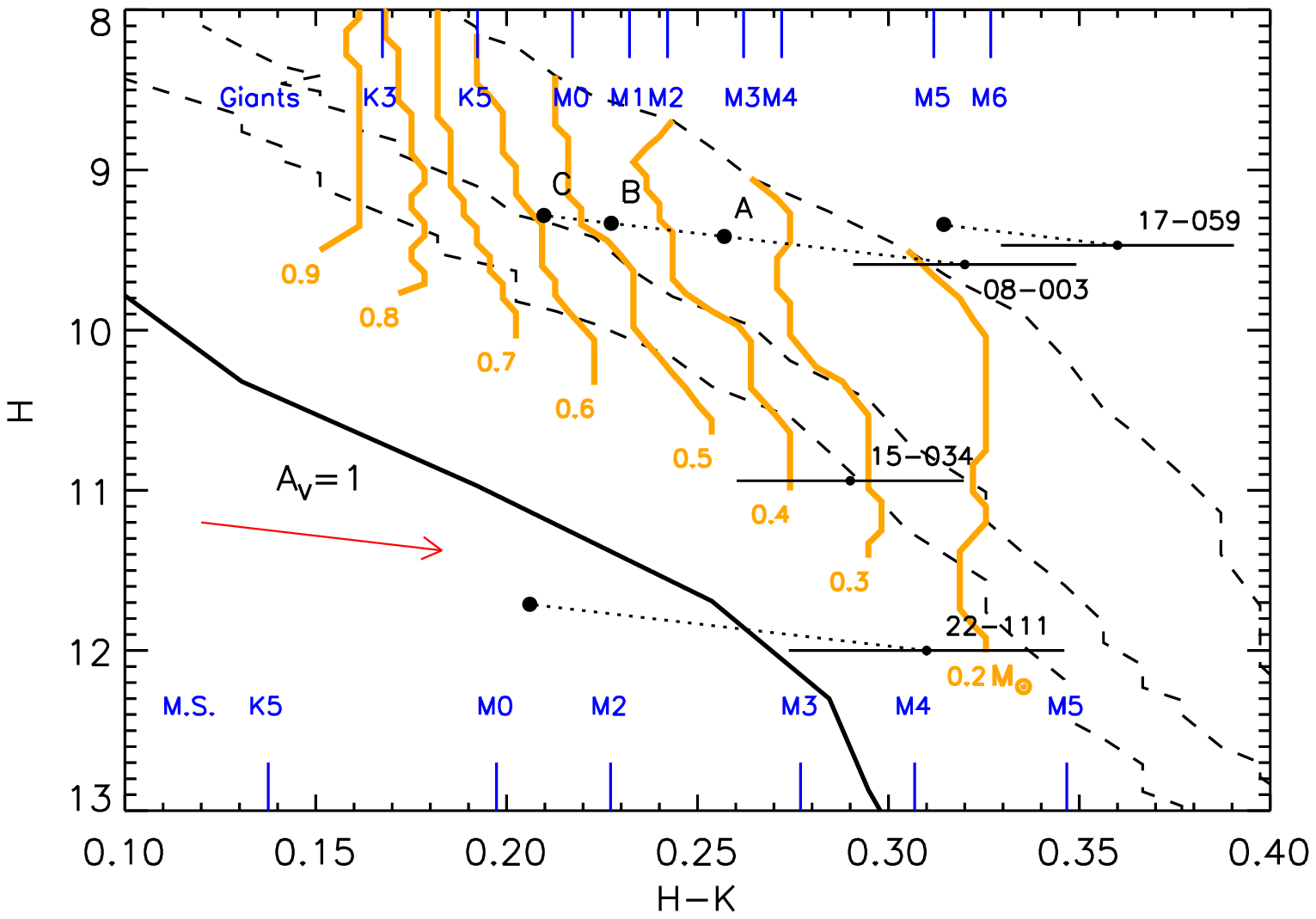}}
\caption{Color-color (\emph{upper panel}) and color-magnitude diagrams
(\emph{middle and lower panels}) illustrating the method used to determine
stellar properties and membership. In each plot, the thin dashed lines are
isochrones at 1, 5, and 10\,Myr; the thick solid line is the MS. Light (yellow
in the online version) solid lines are evolutionary tracks for different masses.
The two vertical dashed lines in the color-color diagram mark the H-K range
expected for an M5 PMS star as \object{XEST-17-059} (see Sect.
\ref{IRdiagr_lrs}). The points A, B and C mark the dereddened positions of the
new TMC member \object{XEST-08-003} on the isochrones in the color-color
diagram, and the positions in the color-magnitude diagrams for relevant 
$A_{\rm V}$ values. The field star \object{XEST-22-111} is shifted back to the
MS in the color-color diagram. \object{XEST-15-034} is an uncertain case.}
\label{fig:stima_massa}
\end{center}
\end{figure}

We shifted the new PMS stars in the color-color diagram along the reddening
vector back to the isochrones at 1, 5, and 10\,Myr (see upper panel of 
Fig. \ref{fig:stima_massa} for the example case of \object{XEST-08-003}), and
then corrected their positions in the color-magnitude diagrams. We visually
assigned the mass, age, and $A_{\rm V}$ values, and relevant ranges, 
that give the best agreement between the three diagrams (middle and lower panels
of Fig. \ref{fig:stima_massa}). 

The weak-lined \object{XEST-09-042} shows relatively low absorption 
($\le 0.3$\,mag), roughly consistent with the 
$N_{\rm H} \sim 10^{21}$\,cm$^{-2}$ measured from
spectral fitting of the X-ray spectrum. We estimated the mass of this star $M=0.6\pm 0.1$\,M$_{\odot}$ and its age to be $\sim 3$\,Myr, in the range
$1-5$\,Myr.

For the weak-lined \object{XEST-08-003}, rather good agreement between the three
diagrams is found assuming a spectral type $early$-M and 
$A_{\rm V}= 1.5\pm 0.2$, consistent with the X-ray measure of $N_{\rm H}$ 
($2.4\times 10^{21}$\,cm$^{-2}$). The mass is estimated to be 
$0.45$\,M$_{\odot}$ (in the range $0.35-0.65$\,M$_{\odot}$) and the age 
$\sim 5$\,Myr (in the range $3-7$\,Myr). 
We recall that this source is likely a binary star (Sect. \ref{rad_rot_vel}),
hence the IR photometry may include contribution from more than a star.
Therefore, the mass and age of this TMC member remain more uncertain. 

The same procedure described above was applied to the new CTTS member
\object{XEST-26-062}, giving mass $0.25\pm 0.05$\,M$_{\odot}$, age 
$\sim 1 (\lesssim 3)$\,Myr, and $A_{\rm V}= 1.0\pm 0.1$. However, IR excess from
the disk is not accounted for, so the mass of this M-type star may be slightly
greater than the value reported here.

The star \object{XEST-11-035} shows little absorption ($A_{\rm V}\sim 0.1$) from
the color-color diagram, and is most likely an old $\sim $\,K5 main sequence
star ($M\sim 0.7$\,M$_\odot$) located at a distance of $\sim 80$\,pc from us,
with low $L_{\rm X}$ ($\sim 3\times 10^{27}$\,erg\,s$^{-1}$). The K1 IV/III
stars \object{XEST-27-022}, \object{XEST-21-059}, and \object{XEST-04-060} show
visual absorption of $\sim$\, 0.5, 0.5, and 2.8, respectively; they have 
$L_{\rm X}/L_{\rm bol} \sim 0.5-1\times 10^{-5}$.

\subsection{Stars observed with DOLORES}
\label{IRdiagr_lrs}

For the stars observed at low resolution we used the IR diagrams with the
aim of providing further support of their nature as young objects or field
stars, and deriving useful stellar properties as in Sect. \ref{IRdiagr_sarg}. 
We do not consider the new CTTS \object{XEST-11-078} here, since its position in
IR diagrams is probably affected by significant IR excess. Despite the
uncertainties of $\sim 2$ subtypes in the spectral type derived from DOLORES
spectra, we can distinguish M-type stars from G- and K-type stars. This is
sufficient for membership confirmation because we can univocally identify
the intersection between the dereddening vector of an absorbed star
with a given evolutionary model. We divide the whole LRS sample into 3
different groups (6 new TMC members, 3 uncertain cases, and 8 field stars), as
explained in the following.

\subsubsection{New TMC members}
\label{IRdiagr_lrs_membri}

The first group is composed of the 6 M-type stars \object{XEST-08-033},
\object{XEST-08-047}, \object{XEST-08-049}, \object{XEST-11-087},
\object{XEST-16-045}, and \object{XEST-17-059} whose low-resolution
spectra (Sect. \ref{dolores}) indicate gravity lower than MS stars of the same
spectral type. In any case, we first explored the hypothesis that these are
indeed field stars. We can exclude these M-type stars being
giants behind the cloud, because their position in the color-color diagram is
inconsistent with the locus of giants, with the exception of
\object{XEST-08-047}, and because X-ray emitting M-type giants are very unusual 
\citep[see, e.g. ][]{Maggio1990,Hunsch1998}. We can also exclude these
stars being active MS stars, because in this hypothesis dereddening onto the MS
track would give significantly greater visual absorption than allowed by the
small derived photometric parallaxes. 

With the method described below, we found that these six stars have
dereddened positions in the IR diagrams that are compatible with PMS stars. For
a given star belonging to this group, we marked in the color-color diagram the
H-K color range whose extremes are defined by the H-K colors of MS 
\citep[from ][]{Kenyon1995} and giant \citep[from ][]{Bessell1988} stars of the
same spectral type of that star, as classified in Sect. \ref{dolores}. Then, we
dereddened the star position in this diagram searching for ``PMS solutions''
with intrinsic H-K color within the above-mentioned range 
(Fig. \ref{fig:stima_massa}). Note that intrinsic colors of PMS stars
depend on the stellar age, which we do not know at this stage. Also the
J-H color may be a better choice than the H-K color to avoid
overestimating the extinction in the case of optically thick disks. However, we
preferred to use the H-K color because J-H shows variations with the age, for a
given spectral type, greater than the former color, thus providing poor
constraints; moreover, the consistency of the results from IR diagrams and
spectral types form LRS spectroscopy (as well as H$\alpha$ emission level)
indicates that these stars have no significant IR excess, with the possible
exception of \object{XEST-11-087} (see below). With this method, we eventually
found the best agreement between the three diagrams for the $A_{\rm V}$, mass,
and age values reported in Table \ref{tab:stars_lrs_memb_unc}. We estimated
errors on these quantities by taking into account the errors on the IR colors
and the LRS-derived spectral type. We note that the spectral type obtained from
the color-color diagram for XEST-11-087 is somewhat later than what is derived
from the LRS spectrum, which may indicate the presence of significant IR
contribution from a disk.  

Therefore, on the basis of all these results and the H$\alpha$
equivalent widths (Table \ref{tab:stars_lrs_memb_unc}), we consider these stars
as new members of the TMC, and we classify them as weak-lined T~Tauri stars.

\subsubsection{Uncertain cases}

The second group includes the three ``uncertain'' stars \object{XEST-08-014},
\object{XEST-15-034}, and \object{XEST-20-071} that deserve further
investigation. The first two are consistent with being either $\sim 10$\,Myr old
Taurus members or main sequence stars in front of the cloud, if we
conservatively consider both the errors on the LRS-derived spectral type and the
IR data (Fig. \ref{fig:stima_massa}) and that the gravity of old 
($\sim 10$\,Myr) T~Tauri stars is much more similar to that of MS stars than
giants (see the case of XEST-15-034 in Fig. \ref{fig:Na_CaH}). Instead, the
nature of the M-type star XEST-20-071 remains uncertain. Its spectrum shows
signs of having gravity higher than in a giant star. However, consistency
between the three IR diagrams could be found neither as an MS nor a PMS star.
From the color-color diagram, a K spectral type is indicated, which is
inconsistent with the result from the low-resolution spectrum.

\subsubsection{Field stars}

Finally, 8 sources are identified as field stars. \object{XEST-06-041},
\object{XEST-12-012}, \object{XEST-15-075}, \object{XEST-19-002}, 
\object{XEST-22-111}, and \object{XEST-27-084} have gravity typical of
main sequence stars, and we found consistency between the three diagrams
assuming they are indeed MS stars\footnote{Dereddening in the color-color 
diagram onto the 10\,Myr isochrone, under the assumption of high-gravity T~Tauri
star, leads to positions in the color-magnitude diagrams significantly below
this isochrone.} (Fig. \ref{fig:stima_massa}).  Although we were not able to
assess the luminosity class of the $early$-G star \object{XEST-18-059}, its
dereddened position surely falls below the main sequence, so it is excluded as a
PMS object. XEST-18-059 may be a G0/1 MS star $\sim 330$\,pc away, with a mass
slightly greater than the solar value and 
$L_{\rm X}\sim 6\times 10^{29}$\,erg\,s$^{-1}$ 
($L_{\rm X}/L_{\rm bol} \sim 6\times 10^{-5}$). The luminosity class of the
$late$-G star \object{XEST-05-027} is also not clear from its LRS spectrum,
however it is excluded as an MS star because of the incompatibility between the
distance it would have in this case ($\sim 50-60$\,pc) and the high visual
absorption $A_{\rm V} \sim 7.5$\,mag (consistent with the X-ray measure of 
$N_{\rm H} = 1.1\times 10^{22}$\,cm$^{-2}$). If \object{XEST-05-027} were a PMS
star, it would be much younger than 1\,Myr from the IR diagrams, and probably
would have significant IR excess. This source is included in the catalog of the 
\emph{Taurus Spitzer Legacy Project} \citep[Data Release 1, ][]{Padgett2006},
and the SED we derived from \emph{Spitzer} and 2MASS photometry shows no signs
of IR excess from circumstellar material, but it is instead typical of normal
stars. Therefore, \object{XEST-05-027} is most likely a giant (or subgiant)
behind the TMC, having $L_{\rm X}/L_{\rm bol} \sim 2\times 10^{-4}$. 

Table \ref{tab:stars_lrs_memb_unc} summarizes the stellar properties derived in
this work for new TMC members and uncertain sources, while Table
\ref{tab:stars_lrs_fs} gives information about the field stars.

\begin{center}
\begin{table*}[t!] 
\caption{Stellar properties for the new Taurus members and the uncertain sources
observed with DOLORES.}
\begin{center}
\begin{tabular}{lccccccc}\hline\hline
star        & Sp. type$^a$ & H$\alpha$ EW$^a$ & $A_{\rm V}\,^b$ & $N_{\rm H}\,^c$ & $M$ & Age (range) & Evol. stage \\ 
            &  &      (\AA\ )      &      (mag)      & ($10^{21}$\,cm$^{-2}$) & (M$_{\odot}$) & (Myr) &      \\ \hline
\multicolumn{8}{c}{new TMC members} \\
\object{XEST-08-033} & M2 IV    & -11   & $1.8\pm 0.1$ & $4.0\pm 1.0$   & $0.35^{+0.1}_{-0.15}$ & $5\ (1-10)$  & WTTS \\ 
\object{XEST-08-047} & M2 IV    &  -6   & $3.2\pm 0.3$ &                & $0.50^{+0.1}_{-0.05}$ & $7\ (5-10)$   & WTTS \\ 
\object{XEST-08-049} & M2 IV    &  -9   & $1.8\pm 0.2$ & $3.0\pm 0.2$   & $0.50^{+0.05}_{-0.15}$ & $5\ (3-10)$     & WTTS \\ 
\object{XEST-11-078} & K2-4 V    & -16.5 & - &                & - & - & CTTS \\ 
\object{XEST-11-087}$^d$ & M3 IV    &  -6   & $2.0\pm 0.2$ &                & $0.25\pm 0.1$ & $3\ (1-5)$ & WTTS \\ 
\object{XEST-16-045} & M3/4 IV  & -8.5  & $0 (<0.1)$  &                & $0.30\pm 0.1$ & $3\ (1-5)$ & WTTS \\ 
\object{XEST-17-059} & M5 IV    & -16   & $0.6\pm 0.1$ & $1.8\pm 0.3$  & $0.20\pm0.05$ & $1 (<3)$ & WTTS \\ \hline
\multicolumn{8}{c}{uncertain stars} \\
\object{XEST-08-014}$^e$ & M2/3  & -8.5  & $\sim 0$      & $\lesssim 0.2$ & 0.4  & $\sim 10$  & WTTS or MS at $\sim 65$\,pc \\ 
\object{XEST-15-034}$^e$ & M4 V      & -5    & $\sim 0-0.15$ & $\lesssim 0.2$ & 0.35 & $\gtrsim 10$ & WTTS or MS at $\sim 120$\,pc \\ 
\object{XEST-20-071} & M?        & -6    & - & $4.1\pm 0.4$   & - & - & ? \\ \hline
\multicolumn{8}{l}{$^a$ From low-resolution spectra (Sect. \ref{dolores}).} \\
\multicolumn{8}{l}{$^b$ From IR color-color diagram (Sect. \ref{IRdiagr_lrs}).} \\
\multicolumn{8}{l}{$^c$ From X-ray spectral fitting \citep{ScelsiXEST2006}.} \\
\multicolumn{8}{l}{$^d$ Possible IR excess may affect the Av, mass and age
values.} \\
\multicolumn{8}{l}{$^e$ Values of $A_{\rm V}$, $M$ and age refer to the PMS case and are not confirmed.} \\
\end{tabular}
\normalsize
\end{center}
\label{tab:stars_lrs_memb_unc}
\end{table*}
\end{center}

\begin{center}
\begin{table*}[t!] 
\caption{Stellar properties for the confirmed field stars
observed with DOLORES.}
\begin{center}
\begin{tabular}{lccccccc}\hline\hline
star        & Sp. type$^a$ & H$\alpha$ EW$^a$ & $A_{\rm V}\,^b$ & $N_{\rm H}\,^c$ & $M$ & $d$ & Evol. stage \\ 
            &  &      (\AA\ )      &      (mag)      & ($10^{21}$\,cm$^{-2}$) & (M$_{\odot}$) & (pc) &      \\ \hline
\object{XEST-05-027} & G5 IV/III & 2 & $7.5\pm 0.3$ & $11.0\pm 1.5$ & - & - & giant \\
\object{XEST-06-041} & M2/3 V & -1.5 & $0.45^{+0.2}_{-0.3}$ &   & $0.50\pm 0.05$ &  $120\pm 20$ & MS \\ 
\object{XEST-12-012} & M2 V & -5 & $2.2\pm 0.3$ &   & $0.50\pm 0.05$ & $155^{+30}_{-15}$ & MS \\ 
\object{XEST-15-075} & M4 V & -7 & $0.3\pm 0.3$ &   & $0.50\pm 0.05$ & $115\pm 15$ & MS \\ 
\object{XEST-18-059}$^d$ & G0/1 & 2.8 & $3.1^{+0.4}_{-0.7}$ &   & $1.15\pm 0.15$ & $330^{+130}_{-60}$ & MS (or giant ?) \\ 
\object{XEST-19-002} & K4 V & 0 & $3.3\pm 0.7$ &   & $0.8\pm 0.1$ &  $400^{+170}_{-100}$ & MS \\ 
\object{XEST-22-111} & M1/2 V & -3.5 & $1.7\pm 0.4$ &   & $0.55\pm 0.05$ &  $175\pm 30$ & MS \\ 
\object{XEST-27-084} & M2 V & -4.5 & $1.2\pm 0.3$ & $0.7\pm 0.4$  & $0.55\pm 0.05$ &   $150\pm 25$ & MS \\ \hline
\multicolumn{8}{l}{$^a$ From low-resolution spectra (Sect. \ref{dolores}).} \\
\multicolumn{8}{l}{$^b$ From IR color-color diagram (Sect. \ref{IRdiagr_lrs}).} \\
\multicolumn{8}{l}{$^c$ From X-ray spectral fitting \citep{ScelsiXEST2006}.} \\
\multicolumn{8}{l}{$^d$ Assuming main sequence star.} \\
\end{tabular}
\normalsize
\end{center}
\label{tab:stars_lrs_fs}
\end{table*}
\end{center}

\section{Discussion} 
\label{dis}

In this work we employed optical spectroscopy, IR photometry, and X-ray data to 
characterize a sample of 25 sources, previously identified as candidate pre-main
sequence stars. We confirm membership in the Taurus-Auriga star-forming region
for 10 of them, while 12 sources are identified as main sequence or older stars
in front of or behind the cloud. Three sources remain as uncertain cases. 
It is worth noting that we have confirmed membership for 5 candidates out of 9
with a high probability assigned by \citet{ScelsiXEST2006} on the basis of the
X-ray analysis; 3 more are the uncertain cases, and only one such source
(\object{XEST-27-084}) has been identified as an active M-type MS star. The
percentage of new Taurus members with respect to the number of observed stars is
at least 40\%, up to $\sim 50$\% depending on the nature of the uncertain
sources. Similar percentages of confirmed new TMC members have been obtained in
other works previously mentioned (Sect. \ref{intro}), where candidate selection
was based on optical and infrared photometry. Hence, our work indicates that
candidate selections based on X-ray emission and infrared photometry are as
effective as selections based on optical and infrared photometry. We also note
that higher percentages of confirmed members could be obtained if the selection
is made considering only X-ray sources within the $M-L_{\rm X}$ relation
discussed below.

It is also interesting that 5 out of 10 newly discovered TMC members (i.e.
the three lithium stars and 2 members observed at low-resolution) are variable
in X-rays (see Table \ref{tab:lista}). Short-term X-ray variability, often
associated to flares, is a frequent characteristic of active stars
\citep{Guedel2003,Wolk2005,Caramazza2007}. The X-ray variability of pre-main
sequence stars in the Taurus molecular cloud has recently been studied by 
\citet{StelzerXEST2006}, who found that about half of a sample of 122 members
detected in the XEST survey are variable. Our result is therefore consistent
with the percentage of X-ray variable members in Taurus.

As explained in Sects. \ref{sarg} and \ref{IRdiagr_lrs_membri}, we found 
8 new WTTS and 2 new CTTS, which is not surprising considering the initial 
X-ray-based candidate selection. The T~Tauri type has been deduced from
the strength of the H$\alpha$ emission and the shape of the line profile for
the stars observed with SARG. The stars \object{XEST-08-003},
\object{XEST-08-033}, \object{XEST-08-047}, \object{XEST-16-045}, and
\object{XEST-17-059} also appear in the catalog of the 
\emph{Taurus Spitzer Legacy Project} (Data Release 1); the photometry obtained
with the IRAC and MIPS cameras onboard the \emph{Spitzer Space Telescope}, 
together with the 2MASS measurements, gives SEDs typical of stars without
significant IR excess for circumstellar material, and thus confirms their nature
as weak-lined T~Tauri stars. 

The new members have spectral types in the range $late$-K --- M so are low-mass
stars, with masses estimated in the range $\sim 0.2-0.6$\,M$_{\odot}$. These
members fall in the same bins where the mass distribution of the $\sim 150$
known Taurus members (within the same XMM fields) shows its bulk.
However, the 25 sources observed at the TNG and analyzed here are the optically
brightest stars of the initial sample selected by \citet{ScelsiXEST2006}, and
therefore other new members with masses of $\sim 0.1$\,M$_{\odot}$, may be
hidden among the 32 remaining candidates. Indeed, about 20 of them are very-low
mass candidates ($M\lesssim 0.2$\,M$_{\odot}$), and about ten of them are brown
dwarf candidates\footnote{The XEST has detected 8 out of 16 known brown dwarfs
included in the survey.}, which will be studied in the continuation of our
optical follow-up program.

\begin{figure}[t!]
\begin{center}
\scalebox{0.5}{
\includegraphics{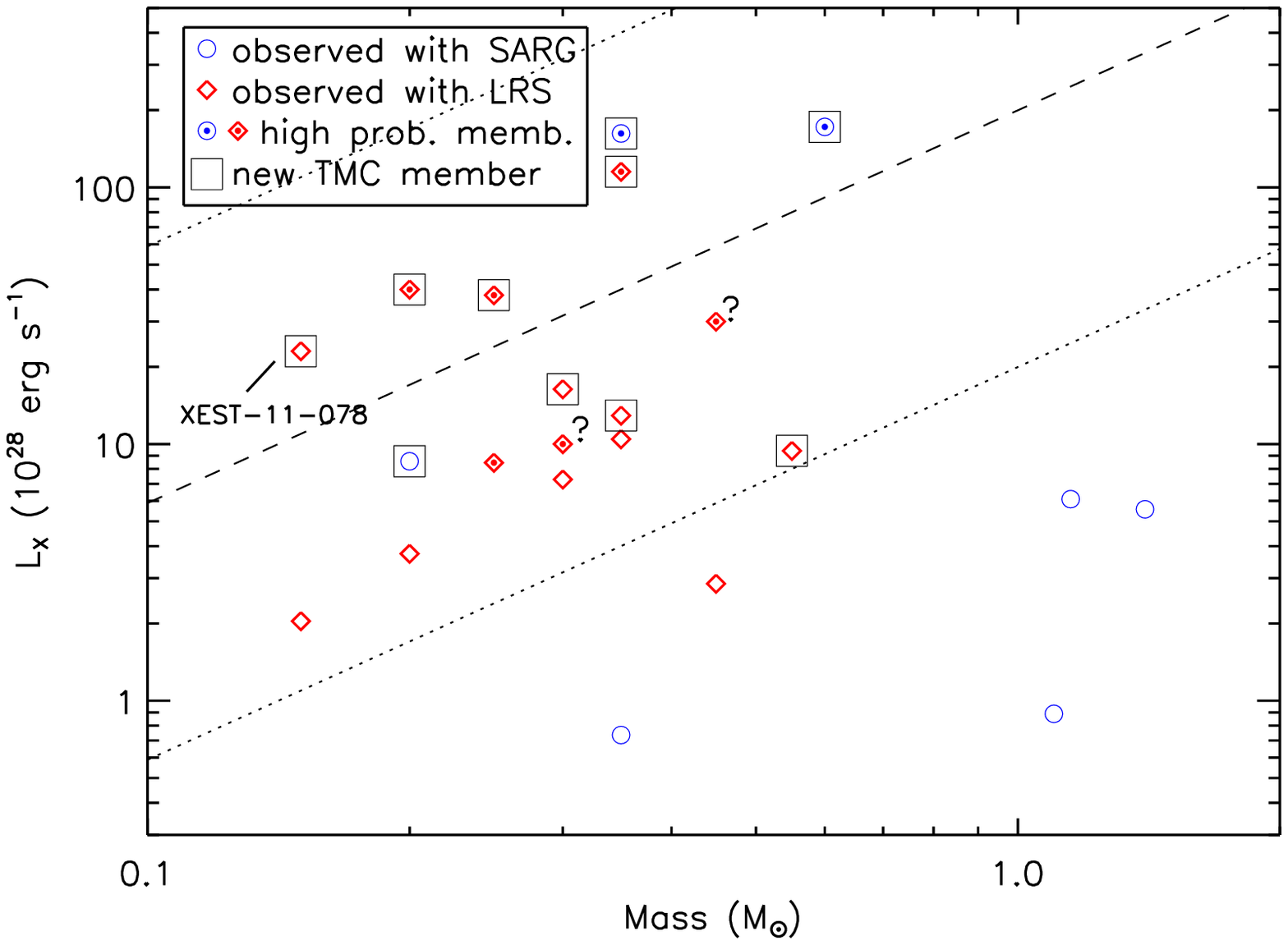}}
\scalebox{0.5}{
\includegraphics{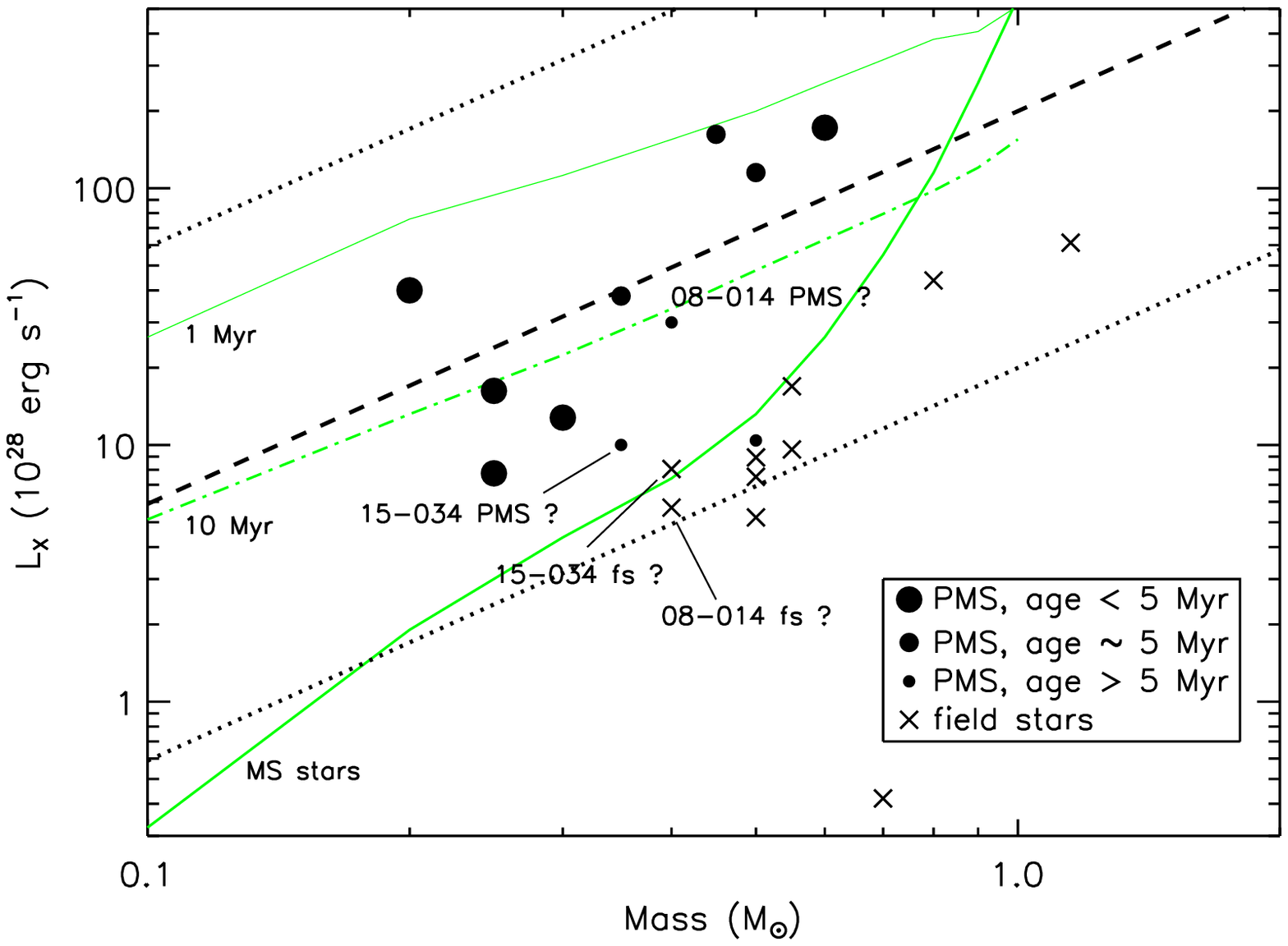}}
\caption{\emph{Upper panel}: Positions of the candidates in the $M-L_{\rm X}$
diagram, with masses and X-ray luminosities estimated before the TNG
observations, assuming they are all members. Circles and diamonds mark stars
observed with SARG and DOLORES, respectively. A dot inside the symbol means the
candidate had a higher probability of membership on the basis of X-ray data. 
Squares indicate new members confirmed from this work, and ``?'' mark uncertain
cases. The dashed line is the best fit to the data for known TMC members and the
dotted lines bracket the region where known members are found. \emph{Lower
panel}: $M-L_{\rm X}$ diagram updated after the TNG observations. Circles mark 
new members from this work, with circle size indicating age, and crosses main
sequence field stars. For the uncertain cases \object{XEST-15-034} and
\object{XEST-08-014} both positions as a PMS or a field star are plotted. Light
green lines are the saturation at 1\,Myr, 10\,Myr, and in the main sequence.}
\label{fig:Lx_M}
\end{center}
\end{figure}
Figure \ref{fig:Lx_M} shows the $M-L_{\rm X}$ diagram for the sources studied in
this work, for which mass estimates have been possible. In the upper panel,
masses and X-ray luminosities were estimated by \citet{ScelsiXEST2006} before
the TNG observations and under the assumption that the candidates were indeed
TMC members\footnote{In this plot, the mass of \object{XEST-11-078} had been
estimated under the wrong assumption of spectral type~M and no IR excess.}. In
this plot we report the relation $\log\,L_{\rm X}=1.54\,\log\,M+30.31$,
best-fitting the data for known TMC members within the fields of the XEST survey
\citep{GuedelXEST2006}, and we also mark the new Taurus members from this study.
All new TMC members are found within the relation derived from known members,
while the five sources outside this relation (i.e. \object{XEST-18-059},
\object{XEST-27-022}, \object{XEST-21-059}, \object{XEST-04-060}, and
\object{XEST-11-035}) are confirmed as non-members. This result both confirms
the validity of a correlation between mass and X-ray luminosity for the young
stars of Taurus and suggests that the $M-L_{\rm X}$ diagram may be used in
future searches for new pre-main sequence stars as a further criterion for the
identification of new candidate members.

The lower panel of the same figure updates the previous plot after the
observations at the TNG and the analysis presented in this paper. In this plot, 
the new members are also distinguished by their age, as calculated in Sects.
\ref{IRdiagr_sarg} and \ref{IRdiagr_lrs}. We did not include giant or subgiants
field stars, since their masses and X-ray luminosities were not estimated. We
marked the saturation curves $L_{\rm X}=10^{-3}\,L_{\rm bol}$ relevant to
different evolutionary stages, i.e. stars at 1\,Myr, at 10\,Myr, and on the main
sequence 
\citep[the relation between bolometric luminosity and mass was taken from the models by][]{Baraffe1998}. Interestingly, the slopes of the theoretical
saturation curves for PMS stars are very similar to the slope of the 
$M-L_{\rm X}$ relation found for Taurus members, suggesting that part of the 
observed spread around this best-fit relation is due to saturated stars at 
different ages. However, both the data in the plot and the narrower $L_{\rm X}$
range spanned by the saturation curves in the $1-10$\,Myr range suggest that
other factors may play a role, such as flares, non-saturated stars (slow
rotators), and the possibly lower $L_{\rm X}$ of CTTS with respect to WTTS
\citep{Preibisch2005,BriggsXEST2006}.

Finally, it is also interesting to note that the saturation curve for main
sequence stars steepens at $M\sim 0.5-0.6$\,M$_{\odot}$, and explains why older
stars may be found within the relation for TMC members. In fact, most
of our confirmed field stars are close to, or in, the saturation regime. 
 
\begin{acknowledgements}

We thank the referee, Dr. J. Alcal{\'a}, for his comments that helped to
improve the quality of the paper. We acknowledge financial contribution from
contract ASI-INAF I/088/06/0. L.S. thanks J. Lopez-Santiago and F. Damiani for
useful discussions, and the contact astronomer at the TNG who performed the
observations in service mode. This publication makes use of data products from
the Two Micron All Sky Survey (2MASS), which is a joint project of the
University of Massachusetts and the Infrared Processing and Analysis
Center/California Institute of Technology, funded by the National Aeronautics
and Space Administration and the National Science Foundation.

\end{acknowledgements}

\bibliographystyle{aa}
\bibliography{sce08}

\end{document}